\documentclass[conference,9pt]{IEEEtran}
\IEEEoverridecommandlockouts
\usepackage{cite}
\usepackage{amsmath,amssymb,amsfonts}
\usepackage{graphicx}
\usepackage{textcomp}
\usepackage{orcidlink}
\usepackage{comment}
\usepackage{stfloats}
\usepackage{setspace}
\usepackage{amssymb}
\usepackage{braket}
\usepackage{xcolor}
\usepackage{placeins}
\usepackage{algorithm}
\usepackage{algorithmic}
\usepackage{graphicx} 
\usepackage{float}
\usepackage{booktabs}
\usepackage{multirow}
\usepackage{adjustbox}
\usepackage{tabularx}
\usepackage[table]{xcolor}
\usepackage[font=footnotesize,skip=0pt]{caption}
\def\BibTeX{{\rm B\kern-.05em{\sc i\kern-.025em b}\kern-.08em
    T\kern-.1667em\lower.7ex\hbox{E}\kern-.125emX}}

\begin{document}

\title{HQTN-SER: Speech Emotion Recognition with Hybrid Quantum Tensor Networks}
\author{
    \IEEEauthorblockN{
        Mahad Mohtashim\textsuperscript{1},
        Nouhaila Innan\textsuperscript{2,3},
        Muhammad Shafique\textsuperscript{2,3}
    }
    \IEEEauthorblockA{
        \textsuperscript{1}School of Electrical Engineering and Computer Science, \\National University of Sciences \& Technology (NUST), Islamabad, Pakistan\\
        \textsuperscript{2}eBRAIN Lab, Division of Engineering, New York University Abu Dhabi (NYUAD), Abu Dhabi, UAE \\
        \textsuperscript{3}Center for Quantum and Topological Systems (CQTS), NYUAD Research Institute, Abu Dhabi, UAE \\
        Emails: mahadm.bscs21seecs@seecs.edu.pk, \{nouhaila.innan, muhammad.shafique\}@nyu.edu
    }
}
\maketitle

\begin{abstract}
Speech emotion recognition (SER) remains fragile in real-world conditions because emotional cues are subtle, speaker-dependent, and easily confounded by recording variability, while high-performing deep models typically rely on large and carefully curated training sets. Quantum machine learning offers an alternative way to introduce nonlinear correlation modeling with compact modules, yet existing quantum SER studies remain limited, and the impact of circuit structure is not well understood. This paper presents HQTN-SER, a hybrid quantum-classical framework that investigates how quantum tensor network connectivity can support SER under small-qubit settings. HQTN-SER introduces (i) an MPS-inspired quantum tensor network module that enforces structured interactions to model correlations in speech representations with a small number of trainable parameters, and (ii) a fusion strategy that combines quantum measurement features with a learned classical latent embedding for end-to-end emotion classification. We evaluate HQTN-SER on three public benchmarks (RAVDESS, SAVEE, and MDER) under a unified preprocessing and training protocol. The proposed model achieves test accuracies of 80.09\% on RAVDESS, 78.12\% on SAVEE, and 73.51\% on MDER, with corresponding macro-F1 scores of 0.8012, 0.7826, and 0.7351, respectively. The model also exhibits stable convergence while using only three or four qubits, suggesting that MPS-inspired circuit structure can be an effective and hardware-aware design choice for quantum-assisted SER. These results provide a compact experimental baseline for studying structured quantum modules in affective computing.

\end{abstract}

\begin{IEEEkeywords}
Quantum Machine Learning, Quantum Neural Networks, Speech Emotion Recognition
\end{IEEEkeywords}
\section{Introduction}

Speech emotion recognition (SER) aims to infer a speaker's affective state from acoustic signals and is increasingly central to human-centered artificial intelligence. It supports applications such as socially aware human-computer interaction, assistive technologies, and mental health monitoring \cite{911197}. Unlike lexical speech recognition, SER depends on subtle and context-sensitive cues in prosody, pitch dynamics, and energy patterns. These cues vary across speakers and recording setups and are easily affected by channel noise and dataset bias, making robust SER inherently challenging \cite{ELAYADI2011572}.

Modern deep learning has pushed SER performance forward, yet the strongest results are typically obtained with high-capacity architectures. Recent CNN-RNN hybrids, attention-based models, and Transformer-style encoders can achieve high accuracy on popular benchmarks, but often with large parameter counts and substantial training and inference costs \cite{article}. This reliance on scale is misaligned with the realities of SER data: widely used datasets are often small, speaker-limited, and class-imbalanced, and they rarely capture the diversity encountered in deployment. As a result, increasing model size can improve in-domain metrics while still leaving models sensitive to distribution shift across speakers, languages, and recording conditions \cite{alhussein2025speech}. This motivates SER pipelines that remain accurate and stable while operating in a parameter-efficient regime.

Classical efficiency-focused designs attempt to reduce cost through lightweight CNNs, shallow networks, and attention-reduced modules \cite{chowdhury2025speech,wu2025comprehensive}. However, parameter reduction alone does not address a key difficulty in SER: many emotions are separated by fine-grained acoustic differences, and reliable recognition requires modeling structured interactions between correlated cues, such as pitch trajectories, spectral tilt, and energy dynamics. When the model budget is tight and data are limited, learning these interactions becomes difficult, and confusion between closely related classes, such as neutral versus calm or sad versus fearful, remains common. The core question is therefore not only how to shrink models, but how to preserve correlation modeling capability when parameters and data are constrained.

Quantum machine learning (QML) provides a different mechanism for constructing compact nonlinear transformations, particularly in the NISQ era, where circuit depth and qubit counts are restricted. Variational quantum circuits (VQCs) can implement nonlinear feature maps through entanglement and interference, offering a structured way to enrich feature interactions with relatively few trainable parameters \cite{Schuld_2019,biamonte2017quantum}. QML has been applied to diverse learning problems, ranging from image classification and time-series analysis to speech-related pipelines and broader pattern recognition tasks \cite{innan2024financial,innan2024quantum,innan2025lep,innan2025quav,choudhary2026hqnn,hong2025review}. For SER, early studies indicate that small circuits can be trained end-to-end, but reported gains are inconsistent and often dataset-specific, and many approaches adopt generic circuit topologies with limited inductive bias \cite{balachandran2025advanced,norval2025quantum,kucharski2025survey}. This makes it hard to separate architectural effects from sensitivity to ansatz choice and tuning \cite{larocca2025barrenplateausvariationalquantum,innan2024financial1,vyskubov2026scaling}.

These observations point to a sharper design question: can structured quantum architectures provide stable and resource-aware behavior for SER under small-qubit and shallow-depth constraints? Tensor network quantum circuits inspired by matrix product states (MPS) offer a principled structure for this setting. MPS connectivity controls parameter growth, restricts entanglement to local neighborhoods, and supports hierarchical correlation modeling \cite{kard}. Such constraints are well matched to speech representations, where information is correlated but not uniformly dependent across all factors. At the same time, existing SER studies rarely use tensor network connectivity, and the practical impact of this structure has not been systematically examined across datasets under a unified protocol.

Motivated by this gap, the goal of this work is not to demonstrate unconditional quantum advantage, but to investigate whether imposing structure in the quantum component can lead to more stable and parameter-efficient SER models under realistic constraints.

To address this gap, we propose HQTN-SER, a hybrid quantum-classical framework that investigates the role of quantum tensor network structure in speech emotion recognition. The framework combines a lightweight classical encoder with an MPS-based quantum module, allowing the model to integrate learned acoustic embeddings with quantum measurement features. The contributions are:
\begin{itemize}
    \item \textbf{Hybrid quantum tensor network module:} We design a quantum processing block with MPS to model structured correlations in speech features using a compact trainable parameterization.
    \item \textbf{Quantum-classical fusion design:} We introduce an end-to-end differentiable fusion strategy that combines quantum measurement statistics with a learned classical latent embedding to improve separability while preserving stable optimization.
    \item \textbf{Multi-dataset evaluation:} We evaluate HQTN-SER on three SER benchmarks, RAVDESS, MDER, and SAVEE, to assess behavior across different speaker distributions, languages, and recording characteristics.
    \item \textbf{Comparative, stability, and hardware-aware analysis:} We benchmark HQTN-SER against prior hybrid quantum SER models under comparable settings and analyze convergence behavior, parameter efficiency, class-wise stability, and hardware-aware robustness using \texttt{FakeMarrakesh}.
\end{itemize}

\section{Related Work}

\subsection{Classical SER: features and compact classifiers}
Early SER pipelines were built around handcrafted acoustic descriptors paired with lightweight classifiers. Typical features include Mel-frequency cepstral coefficients (MFCCs), prosodic cues such as pitch and energy, spectral statistics, and voice-quality measures. These representations are commonly combined with support vector machines, Gaussian mixture models, random forests, or shallow neural networks, and they remain attractive for data-limited settings due to their modest parameter counts and stable optimization behavior.

As computational resources and annotated data became more accessible, deep learning architectures became dominant in SER \cite{swain2018databases,khalil2019speech,issa2020speech}. Convolutional neural networks (CNNs) applied to MFCCs or time-frequency representations treat SER as a pattern recognition problem \cite{mao}, while recurrent models (RNNs, LSTMs) explicitly model temporal structure in speech \cite{solanki2025evaluating}. More recently, attention-based and Transformer architectures have been introduced to capture longer-range dependencies and contextual cues that can affect perceived emotion \cite{sperber2019selfattentionalmodelslatticeinputs}. In parallel, large-scale self-supervised and foundation speech models further demonstrate the effectiveness of representation learning at scale. For example, encoder-decoder Transformer models trained on massive multilingual corpora can provide robust features for diverse speech tasks\cite{radford2022robustspeechrecognitionlargescale}.

However, high accuracy in these settings often comes with substantial computational and parameter cost. This has motivated compact SER architectures designed to reduce inference overhead while preserving accuracy, including lightweight CNN variants, gated recurrent units (GRUs), attention-reduced Transformers, and hybrid CNN--RNN designs \cite{latif}. Despite these advances, compact classical models still face difficulty when emotion categories are separated by fine-grained acoustic differences \cite{barradas2026emotion}. In particular, reliable recognition often requires modeling structured interactions among correlated cues rather than relying on isolated descriptors, and learning such interactions under tight parameter budgets remains challenging. This motivates exploring architectures that impose correlation structure by design, rather than increasing model size.

\subsection{QML for audio and SER}
QML has been investigated for compact nonlinear feature transformations for audio and speech-related tasks. Early studies mainly focused on quantum kernels and VQCs applied to small-scale classification, demonstrating that audio-derived features can be embedded into quantum states and used in learning pipelines. Subsequent work explored hybrid quantum-classical models where a classical front-end extracts acoustic features and a parameterized quantum circuit performs a learned transformation before a classical prediction head.

In SER, Soltani et al. \cite{soltani} proposed a quantum-enhanced echo-state-network-based approach, reporting competitive accuracy and favorable convergence behavior on benchmark datasets. Mittal et al. \cite{mittal2025hybrid} introduced a hybrid framework combining classical preprocessing with parameterized quantum circuits for emotion classification, suggesting that quantum components can provide useful nonlinear mappings under constrained parameter budgets. Related directions incorporate privacy and distributed training objectives, along with broader efforts on representation learning, robust model design, and domain-specific QML pipelines \cite{qu,rajapakshe2025representation,dave2025sentiqnf,krishna2024gesture,golchha2025quantum}.

While promising, much of the existing quantum SER literature differs in circuit choices, preprocessing, and evaluation protocols, which makes comparisons difficult and often leaves the source of performance gains unclear. A recurring limitation is that many models rely on a generic or shallow ansatz, making results sensitive to hyperparameters and offering limited architectural inductive bias for correlated speech representations.

\subsection{Quantum tensor networks and MPS quantum circuits}
Tensor networks provide a structured representation of high-dimensional states through low-rank factorizations, with MPS being a widely studied family \cite{Or_s_2014}. An MPS describes a global state via a chain of local tensors, where a bond-dimension parameter controls the amount of correlation and entanglement that can be represented between different parts of the system. This yields an explicit mechanism to control expressivity and parameter growth, which is attractive for learning under limited resources.

In QML, MPS quantum circuits have gained attention because they enforce local connectivity and regulate entanglement\cite{kard}. Beyond quantum circuits, tensor network-based models have also shown strong results for classical sequence modeling and image classification with significantly fewer parameters than many deep architectures \cite{NIPS2016_5314b967}.

Despite these developments, the use of quantum tensor network structure for SER remains limited. Existing quantum SER pipelines rarely exploit structured quantum connectivity tailored to correlated speech representations, and the benefits of MPS-style inductive bias have not been systematically evaluated across multiple SER datasets under a unified protocol. This motivates the design and analysis of hybrid SER models where the quantum component is structured by construction.

\section{Methodology}
\begin{figure*}[htpb]
    \centering
\includegraphics[width=1\linewidth]{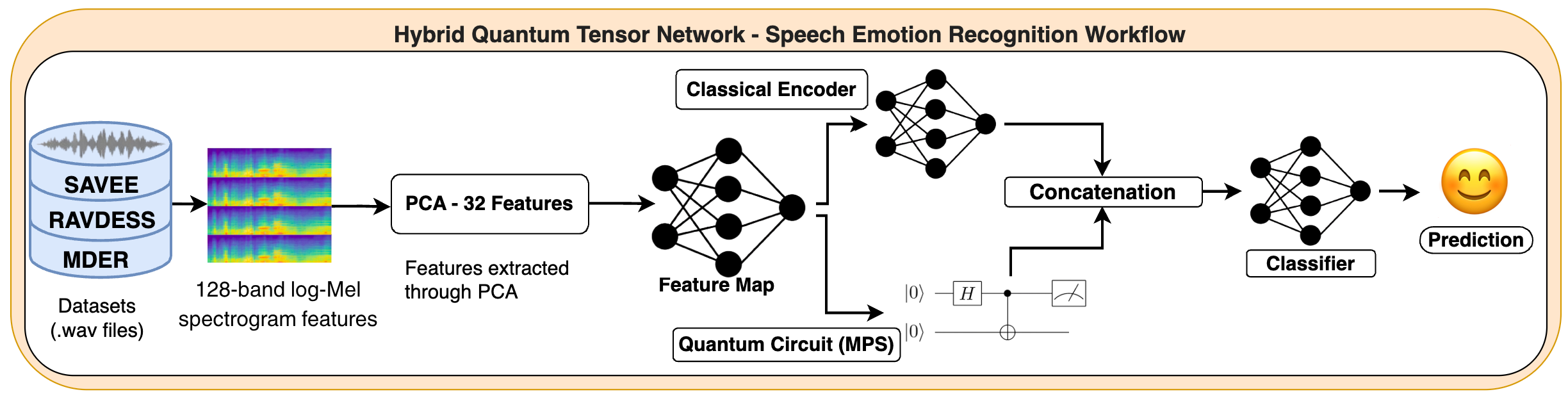}
    \caption{Raw audio files from SAVEE, RAVDESS, and MDER are converted into fixed-length 128-band log-Mel spectrogram representations using Librosa. The vectorized representations are reduced to 32 dimensions using PCA fitted on the training partition only.  The resulting feature vector is mapped into a shared representation and processed in parallel by a classical encoder and an MPS-structured variational quantum circuit; their outputs are concatenated and fed to a final classifier to produce the emotion prediction.}
    \label{fig:hqtn}
\end{figure*}

This section presents HQTN-SER, a hybrid quantum-classical framework for speech emotion recognition that combines a compact classical embedding with a structured quantum tensor network module. The pipeline is designed to keep the quantum component small and structured: acoustic features are standardized into fixed-length time-frequency representations, compressed to a low-dimensional vector, and then processed in parallel by (i) a classical encoder that learns a latent embedding and (ii) an MPS-structured variational quantum circuit that produces measurement statistics. These two representations are fused to produce the final emotion prediction. Fig.~\ref{fig:hqtn} provides an overview of the workflow.

\subsection{Problem definition}
Let $\mathcal{D}=\{(\mathbf{x}_i,y_i)\}_{i=1}^{N}$ denote a labeled SER dataset, where $\mathbf{x}_i \in \mathbb{R}^{T \times F}$ represents an utterance with $T$ time frames and $F$ acoustic features, and $y_i \in \{1,\dots,C\}$ is the corresponding emotion label among $C$ classes. The objective is to learn a predictor $f_{\Theta}$ that maps an utterance to a label, $f_{\Theta}:\mathbf{x}\mapsto \hat{y}$. In HQTN-SER, prediction is obtained from a fused representation that combines a learned classical embedding $\mathbf{z}_c$ with quantum measurement statistics $\mathbf{z}_q$:
\begin{equation}
\hat{\mathbf{y}} = h_{\phi}\big([\mathbf{z}_c \,\|\, \mathbf{z}_q]\big),
\end{equation}
where $[\cdot \,\|\, \cdot]$ denotes concatenation and $h_{\phi}(\cdot)$ is the final classifier. The full set of trainable parameters is denoted by $\Theta$, including the classical parameters and the quantum circuit parameters.
\subsection{Data preprocessing}
\subsubsection{Acoustic feature extraction and standardization}
Each audio signal is converted into a fixed-length time-frequency representation so that inputs are comparable across datasets and compatible with the quantum module. Given a waveform $s(t)$, we resample to $22{,}050$ Hz and truncate or zero-pad each clip to a maximum duration of $5$ seconds. We compute a Mel-spectrogram with $n_{\text{mels}}=128$:
\begin{equation}
\mathbf{M} = \mathrm{MelSpec}(s(t); n_{\text{mels}}=128)\in\mathbb{R}^{T\times 128},
\end{equation}
then convert power values to the decibel scale,
\begin{equation}
\mathbf{M}_{\mathrm{dB}} = 10\log_{10}\!\left(\frac{\mathbf{M}}{\max(\mathbf{M})}\right).
\end{equation}
To enforce a uniform temporal dimension, the time axis is truncated or zero-padded to a fixed length $T_{\max}$, yielding $\mathbf{M}'\in\mathbb{R}^{T_{\max}\times 128}$. Finally, the representation is vectorized:
\begin{equation}
\tilde{\mathbf{x}}=\mathrm{vec}(\mathbf{M}')\in\mathbb{R}^{D}, \quad D=T_{\max}\cdot 128.
\end{equation}

\subsubsection{Dimensionality compression and quantum input mapping}
The vectorized representation is high-dimensional, while the quantum circuit operates on a small number of qubits. We apply principal component analysis (PCA) to compress features while preserving dominant variance. Let $\mathbf{X}\in\mathbb{R}^{N\times D}$ denote the matrix of vectorized features. PCA projects each sample to $k$ dimensions:
\begin{equation}
\mathbf{x}=\tilde{\mathbf{x}}\mathbf{W}_k \in \mathbb{R}^{k},
\end{equation}
where $\mathbf{W}_k\in\mathbb{R}^{D\times k}$ contains the top $k$ eigenvectors of the covariance matrix. We set $k=32$.

Since the quantum circuit uses $n$ qubits ($n\in\{3,4\}$ in our experiments), we map the PCA vector to an $n$-dimensional input through a learnable affine projection:
\begin{equation}
\mathbf{u}=\mathbf{P}\mathbf{x}+\mathbf{b}\in\mathbb{R}^{n},
\end{equation}
where $\mathbf{P}\in\mathbb{R}^{n\times k}$ and $\mathbf{b}\in\mathbb{R}^{n}$ are trained end-to-end. This step resolves the dimensionality mismatch and allows the model to learn which compressed acoustic directions are most informative for the quantum transformation.

Emotion labels are mapped to integer indices $y\in\{0,\dots,C-1\}$. For datasets with available speaker information, we store a speaker identifier $g \in \{0,\ldots,S-1\}$, which can be used to support speaker-independent evaluation and avoid speaker leakage.

\subsection{Classical latent embedding}
A compact encoder transforms the PCA feature vector into a latent embedding:
\begin{equation}
\mathbf{z}_c=g_c(\mathbf{x})\in\mathbb{R}^{d_c},
\end{equation}
where $d_c$ is the embedding dimension. This embedding is trained jointly with the quantum module and later fused with quantum measurement statistics for classification.

\subsection{QTN with MPS connectivity}
\subsubsection{Angle encoding}
Given $\mathbf{u}=[u_1,\dots,u_n]\in\mathbb{R}^{n}$, the quantum input state is prepared using angle encoding:
\begin{equation}
|\psi(\mathbf{u})\rangle = U_{\mathrm{enc}}(\mathbf{u})|0\rangle^{\otimes n}, \quad
U_{\mathrm{enc}}(\mathbf{u})=\bigotimes_{k=1}^{n} R_y(u_k).
\end{equation}

\subsubsection{MPS-structured variational circuit}
\begin{figure*}[htpbt]
    \centering
    \includegraphics[width=\textwidth]{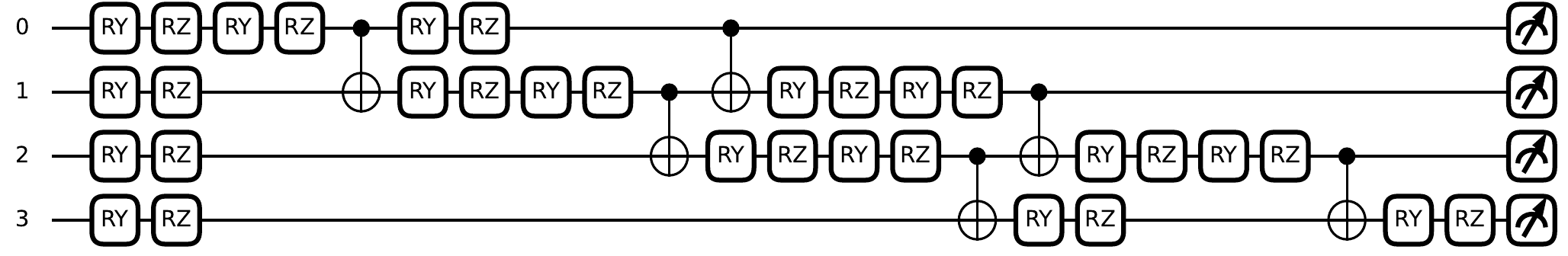}
    \caption{MPS-structured variational circuit used in HQTN-SER. Trainable single-qubit rotations ($R_y$, $R_z$) are interleaved with nearest-neighbor CNOT gates in a left-to-right sweep. Measurement statistics are obtained as expectation values on each qubit and used as quantum features.}
    \label{fig:mps_circuit}
\end{figure*}

The variational unitary $U_{\mathrm{QTN}}(\boldsymbol{\theta})$ follows a matrix-product-state (MPS) pattern with strictly nearest-neighbor entanglement arranged in a left-to-right sweep, matching Fig.~\ref{fig:mps_circuit}. Local trainable blocks are placed on each qubit and interleaved with CNOT gates between adjacent qubits only. The state after the quantum transformation is
\begin{equation}
|\Psi(\mathbf{u};\boldsymbol{\theta})\rangle = U_{\mathrm{QTN}}(\boldsymbol{\theta})\,|\psi(\mathbf{u})\rangle.
\end{equation}
Each local block is parameterized as
\begin{equation}
U_k(\boldsymbol{\theta}_k)=R_y(\theta_k^{(y)})\,R_z(\theta_k^{(z)}),
\end{equation}
and entanglement is introduced only through nearest-neighbor CNOTs along the 1D chain. This connectivity constrains interactions to local neighborhoods, regulates parameter growth, and provides a structured mechanism for modeling correlated acoustic cues.

\subsubsection{Measurement statistics}
After applying $U_{\mathrm{QTN}}(\boldsymbol{\theta})$, quantum features are extracted as expectation values of single-qubit observables:
\begin{equation}
z_{q,k}=\langle \Psi(\mathbf{u};\boldsymbol{\theta})| Z^{(k)} |\Psi(\mathbf{u};\boldsymbol{\theta})\rangle,\quad k=1,\dots,n,
\end{equation}
and collected into $\mathbf{z}_q=[z_{q,1},\dots,z_{q,n}]\in\mathbb{R}^{n}.$
These measurement statistics are differentiable with respect to both the circuit parameters $\boldsymbol{\theta}$ and the projection parameters $(\mathbf{P},\mathbf{b})$.

\subsection{Fusion and emotion classification}
The classical embedding and quantum measurement statistics are concatenated, $\mathbf{z}_f=[\mathbf{z}_c \,\|\, \mathbf{z}_q],$
and passed to a fully connected classifier that outputs class probabilities:
\begin{equation}
\hat{\mathbf{y}}=\mathrm{softmax}(\mathbf{W}_f\mathbf{z}_f+\mathbf{b}_f).
\end{equation}
Concatenation is used as a minimal fusion mechanism that preserves information from both representations without imposing additional structural assumptions.

\subsection{Training objective and optimization}
HQTN-SER is trained end-to-end using categorical cross-entropy:
\begin{equation}
\mathcal{L}=-\frac{1}{B}\sum_{i=1}^{B}\sum_{c=1}^{C} y_{i,c}\log \hat{y}_{i,c},
\end{equation}
where $y_{i,c}$ is the one-hot label, $B$ is the mini-batch size, and $\hat{y}_{i,c}$ is the predicted probability for class $c$. Gradients with respect to the quantum parameters $\boldsymbol{\theta}$ are computed via the parameter-shift rule, enabling joint optimization with classical backpropagation. Optimization uses AdamW with early stopping based on validation performance.

\section{Results and Discussion}
\subsection{Experimental Setup}
We evaluate HQTN-SER on three public speech emotion recognition benchmarks selected to probe robustness under differences in language, speaker composition, recording conditions, and class balance.

\textbf{RAVDESS (Ryerson Audio-Visual Database of Emotional Speech and Song)} \cite{ravdess} contains recordings from 24 professional actors (12 male, 12 female) and covers eight emotions.

\textbf{SAVEE (Surrey Audio-Visual Expressed Emotion)} \cite{haq} is a controlled benchmark recorded from four male speakers across seven emotions.

\textbf{MDER (Moroccan Dialect Emotion Recognition Dataset)} \cite{mderr} focuses on Moroccan Arabic (Darija) and includes natural speech reflecting dialect-specific prosody and speaking style.
\begin{table}[htpb]
\centering
\caption{Emotion distribution of speech samples in the evaluated datasets.}
\label{tab:emotion-distribution}
\small

\begin{tabularx}{\columnwidth}{@{}l *{3}{>{\raggedleft\arraybackslash}X}@{}}
\hline
\textbf{Emotion} & \textbf{RAVDESS} & \textbf{SAVEE} & \textbf{MDER} \\
\hline
Angry     & 192  & 60  & 400 \\
Calm      & 192  & --  & --  \\
Disgust   & 192  & 60  & --  \\
Fearful   & 192  & 60  & 400 \\
Happy     & 192  & 60  & 400 \\
Neutral   & 96   & 120 & 400 \\
Sad       & 192  & 60  & 400 \\
Surprise  & 192  & 60  & --  \\
\hline
\textbf{Total} & \textbf{1440} & \textbf{480} & \textbf{2000} \\
\hline
\end{tabularx}
\end{table}

All experiments are conducted using the train/validation/test splits presented in Table~\ref{tab:hyperparams}. For RAVDESS, the partitions are speaker-disjoint to avoid speaker leakage. SAVEE and MDER are split at the sample level. Preprocessing is performed without leakage: normalization and PCA are fitted on the training split only and then applied unchanged to the validation and test splits.

Performance is reported using per-class precision, recall, and F1-score, with their macro-averages. Unless otherwise stated, all model variants (HQTN-SER and ablations) are trained under the same optimizer and training budget to enable controlled comparison. Quantum circuits are executed using the PennyLane Lightning qubit simulator (\texttt{lightning.qubit}). 


\begin{table}[htpb]
\centering
\caption{Training hyperparameters used across RAVDESS, SAVEE, and MDER datasets.}
\label{tab:hyperparams}
\small
\begin{tabular}{l c c c}
\hline
\textbf{Parameter} & \textbf{RAVDESS} & \textbf{SAVEE} & \textbf{MDER} \\
\hline
Train/Val/Test split & 70/15/15 & 60/20/20 & 60/20/20 \\
Max epochs & 50 & 75 & 40 \\
Qubits ($n$) & 4 & 3 & 3 \\
MPS layers ($L$) & 2 & 1 & 1 \\
Batch size & 16 & 8 & 8 \\
Learning rate ($LR_{MPS}$) & 0.1 & 0.05 & 0.05 \\
Learning rate ($LR_{Classic}$) & $10^{-3}$ & $10^{-3}$ & $10^{-3}$ \\
\hline
\multicolumn{4}{c}{\textit{Common Parameters}} \\
Sampling rate & \multicolumn{3}{c}{22,050 Hz} \\
Mel bands & \multicolumn{3}{c}{128} \\
Max audio duration & \multicolumn{3}{c}{5 s} \\
PCA Components ($k$) & \multicolumn{3}{c}{32} \\
Optimization & \multicolumn{3}{c}{AdamW} \\
\hline
\end{tabular}
\end{table}

\subsection{Convergence Analysis}
\begin{figure}[htpb]
    \centering
    \includegraphics[width=\linewidth]{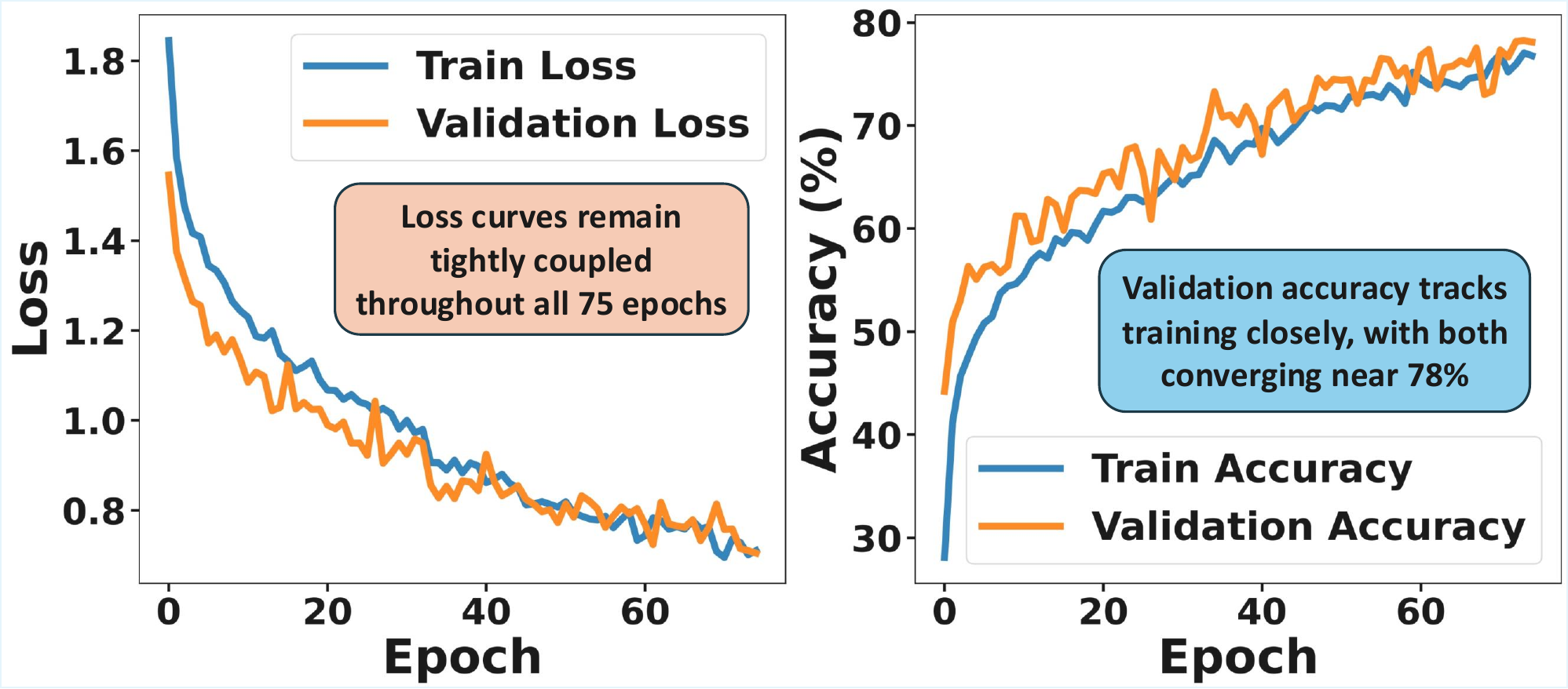}
    \caption{Training and validation loss/accuracy curves on SAVEE.}
    \label{fig:savee_conv}
\end{figure}
To assess optimization stability, we track training and validation loss and accuracy over epochs for each dataset. On SAVEE (Fig.~\ref{fig:savee_conv}), loss decreases steadily, and accuracy increases smoothly toward a stable plateau, with closely aligned training and validation trends and no late-epoch separation, indicating controlled generalization.
\begin{figure}[htpb]
    \centering
    \includegraphics[width=\linewidth]{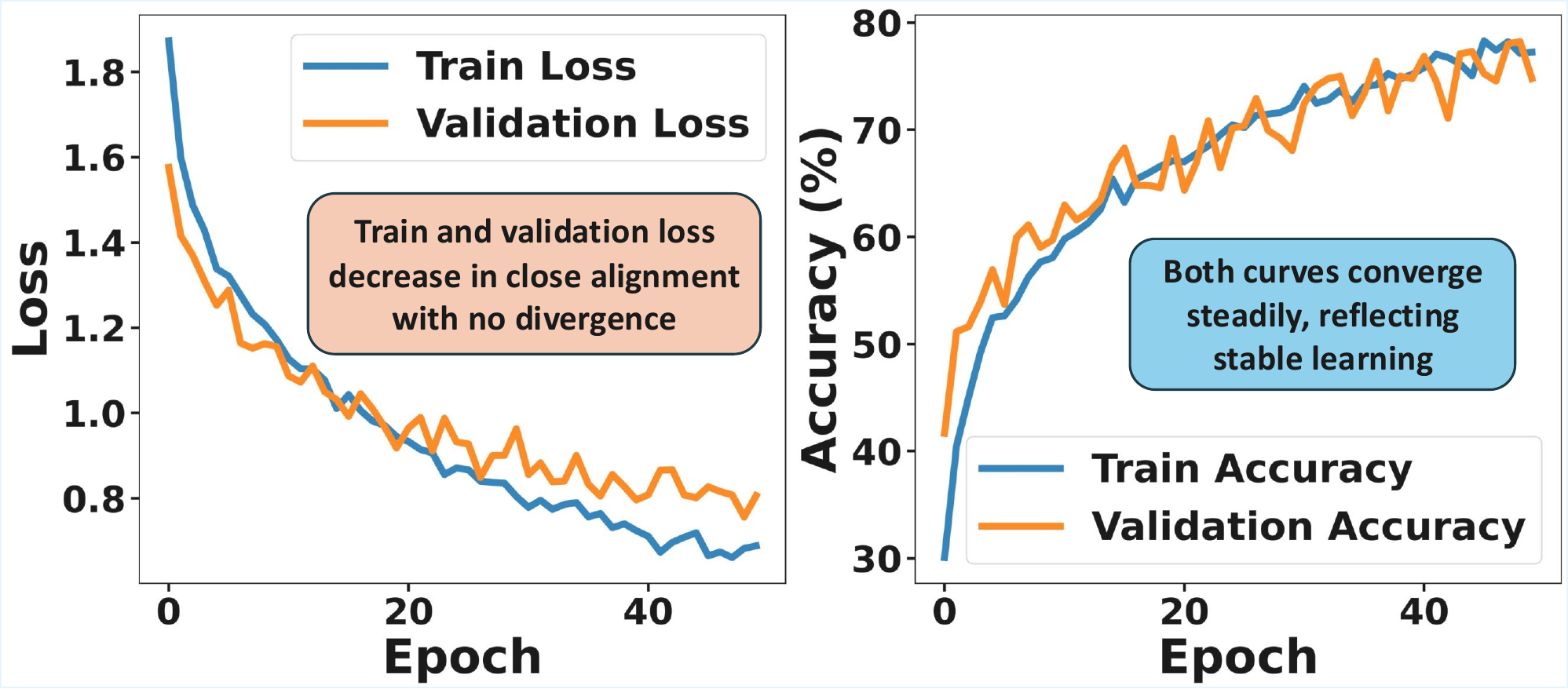}
    \caption{Training and validation loss/accuracy curves on RAVDESS.}
    \label{fig:ravdess_conv}
\end{figure}
On RAVDESS (Fig.~\ref{fig:ravdess_conv}), both losses exhibit a consistent downward trend while validation accuracy shows mild fluctuations expected from a more diverse multi-class setting; however, the training and validation curves remain close throughout training, suggesting stable learning dynamics. 
\begin{figure}[htpb]
    \centering
    \includegraphics[width=\linewidth]{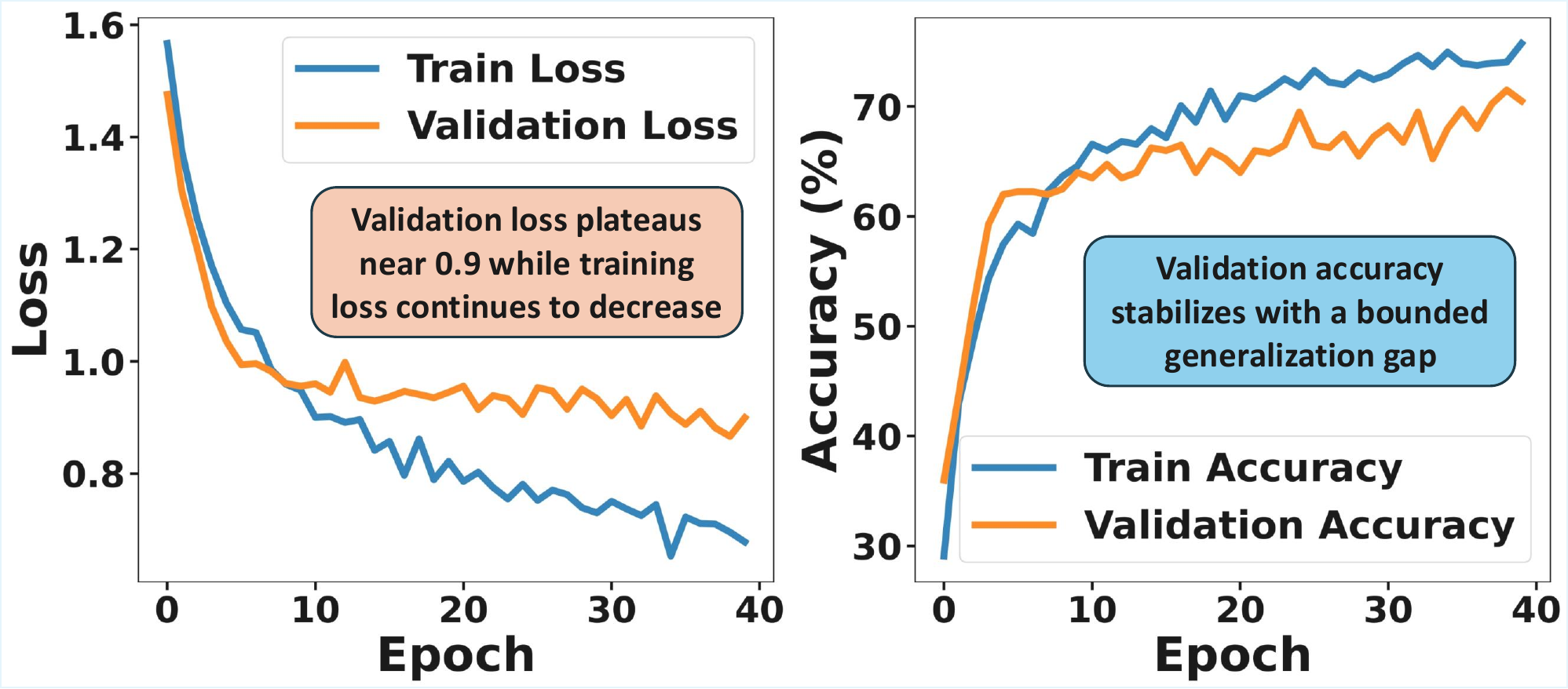}
    \caption{Training and validation loss/accuracy curves on MDER.}
    \label{fig:mder_conv}
\end{figure}
On MDER (Fig.~\ref{fig:mder_conv}), the model learns rapidly in the first epochs and then transitions to a stable regime with a more visible but bounded train--validation gap: training loss continues to decrease while validation loss saturates and validation accuracy improves more gradually. Importantly, none of the datasets exhibit late-stage divergence or oscillatory behavior, supporting that the hybrid optimization remains well-behaved across different languages and recording conditions.
\subsection{Performance Analysis}
To quantify recognition quality beyond a single aggregate score, we report per-class precision, recall, and F1, with macro-averaged performance for each dataset, as shown in Table~\ref{tab:perclass_all}. On SAVEE, HQTN-SER achieves balanced performance across emotions, with the strongest results for \emph{Neutral} (F1 $=0.8363$) and consistently high scores for the remaining classes (F1 $\approx 0.74$--$0.78$), yielding a macro-F1 of $0.7826$. On RAVDESS, the model attains the highest macro-averaged performance among the evaluated datasets (macro-F1 $=0.8012$), with particularly strong recognition of \emph{Surprise} (F1 $=0.8807$) and \emph{Angry} (F1 $=0.8448$). The most challenging category is \emph{Neutral} (F1 $=0.6415$), which is consistent with its acoustic proximity to low-arousal states and its tendency to overlap with other emotions in prosodic space. On MDER, the macro-F1 reaches $0.7351$, with relatively compact variation across classes (F1 $=0.7160$--$0.7578$), indicating stable behavior under a dialectal setting but also reflecting increased linguistic variability compared to the English benchmarks. Overall, the per-class breakdown indicates that the proposed model remains stable across diverse conditions, while the remaining errors cluster in confusable, low-arousal categories where acoustic boundaries are subtle and strongly overlapping.
\begin{table}[htpb]
\centering
\caption{Per-class performance across datasets.}
\label{tab:perclass_all}
\small
\begin{tabularx}{\columnwidth}{@{}l l *{3}{>{\raggedleft\arraybackslash}X}@{}}
\hline
\textbf{Dataset} & \textbf{Emotion} & \textbf{Precision} & \textbf{Recall} & \textbf{F1} \\
\hline
\multirow{8}{*}{\rotatebox{90}{\textbf{SAVEE}}}
& Angry     & 0.7500 & 0.7500 & 0.7500 \\
& Disgust   & 0.7333 & 0.8333 & 0.7805 \\
& Fearful      & 0.7857 & 0.7333 & 0.7586 \\
& Happy     & 0.7692 & 0.7142 & 0.7407 \\
& Neutral   & 0.8214 & 0.8518 & 0.8363 \\
& Sad       & 0.7500 & 0.7500 & 0.7500 \\
& Surprise & 0.8182 & 0.7500 & 0.7826 \\
\rowcolor{green!10}
& \textbf{Macro average} & \textbf{0.7841} & \textbf{0.7812} & \textbf{0.7826} \\
\hline

\multirow{9}{*}{\rotatebox{90}{\textbf{RAVDESS}}}
& Neutral   & 0.7083 & 0.5862 & 0.6415 \\
& Calm      & 0.7727 & 0.8947 & 0.8293 \\
& Happy     & 0.6761 & 0.8421 & 0.7500 \\
& Sad       & 0.7273 & 0.6897 & 0.7080 \\
& Angry     & 0.8448 & 0.8448 & 0.8448 \\
& Fearful   & 0.8704 & 0.8103 & 0.8393 \\
& Disgust   & 0.8679 & 0.8070 & 0.8364 \\
& Surprise & 0.9412 & 0.8276 & 0.8807 \\
\rowcolor{green!10}
& \textbf{Macro average} & \textbf{0.8075} & \textbf{0.8009} & \textbf{0.8012} \\
\hline

\multirow{6}{*}{\rotatebox{90}{\textbf{MDER}}}
& Neutral  & 0.7403 & 0.7125 & 0.7261 \\
& Happy    & 0.7073 & 0.7250 & 0.7160 \\
& Sad      & 0.7059 & 0.7500 & 0.7273 \\
& Angry    & 0.7531 & 0.7625 & 0.7578 \\
& Fearful  & 0.7733 & 0.7250 & 0.7484 \\
\rowcolor{green!10}
& \textbf{Macro average} & \textbf{0.7360} & \textbf{0.7350} & \textbf{0.7351} \\
\hline
\end{tabularx}
\end{table}

\subsection{Ablation Analysis}
To quantify the effect of each component, we evaluate three reduced variants alongside the full HQTN-SER model while keeping the same preprocessing, data splits, and all other training hyperparameters. The first variant removes the quantum module and relies only on the classical encoder and classifier (Classical-only). The second, \textit{Matched Classical}, increases the capacity of the classical network to approximately match the parameter count of HQTN-SER while excluding the quantum module, allowing us to determine whether performance gains arise from additional model capacity alone. The third removes the classical encoder and predicts emotions from the quantum measurement features produced by the MPS circuit alone (Quantum-only). The full HQTN-SER model combines the classical latent embedding with the quantum measurement statistics and applies the same prediction head.

Table~\ref{tab:ablation} shows that the Classical-only baseline achieves moderate performance on MDER and RAVDESS but drops substantially on SAVEE, where limited speakers and short utterances make class boundaries more sensitive to representation quality. The Matched Classical variant provides only marginal improvements over the Classical-only model on SAVEE and MDER, while achieving a larger gain on RAVDESS, indicating that increasing the number of classical parameters alone does not fully account for the improvements obtained by the proposed hybrid architecture. The Quantum-only variant underperforms across all datasets, which is expected under small-qubit, shallow-depth settings where the quantum module alone has limited capacity to form a robust decision boundary from acoustic features. In contrast, HQTN-SER yields a large improvement on all benchmarks, reaching $0.7812$ on SAVEE, $0.7351$ on MDER, and $0.8009$ on RAVDESS. These results indicate that the MPS-structured quantum module is most effective when used as a structured feature transformation within a hybrid pipeline, rather than as a standalone classifier, and that its constrained connectivity can still introduce useful correlation structure under tight quantum resource budgets. Finally, the Classical-only variant is a protocol-matched baseline intended to isolate the role of the quantum module under the same preprocessing and training budget, rather than a tuned or pretrained state-of-the-art SER system.

\begin{table}[htpb]
\centering
\caption{Ablation results (accuracy) across datasets.}
\label{tab:ablation}
\small
\begin{tabularx}{\columnwidth}{@{}l *{3}{>{\raggedleft\arraybackslash}X}@{}}
\hline
\textbf{Model variant} & \textbf{SAVEE} & \textbf{MDER} & \textbf{RAVDESS} \\
\hline
Classical-only & 0.4583 & 0.6850 & 0.6644 \\
Matched Classical  & 0.4792 & 0.6875 & 0.7524 \\
Quantum-only   & 0.2917 & 0.5650 & 0.3426 \\
\rowcolor{green!10}
\textbf{HQTN-SER} & \textbf{0.7812} & \textbf{0.7351} & \textbf{0.8009} \\
\hline
\end{tabularx}
\end{table}
\subsection{Comparative Analysis}
We compare HQTN-SER with representative quantum SER methods from prior work using (i) accuracy on the same benchmark datasets and (ii) quantum footprint in terms of qubit count. We additionally report the \emph{total} number of trainable parameters (classical + quantum) as stated in the corresponding papers, summarized in Table~\ref{tab:comparative-results}. On RAVDESS, HQTN-SER reaches $0.8009$ accuracy with 4 qubits, matching or slightly exceeding the listed baselines that typically use larger quantum registers (5--10 qubits). On SAVEE, HQTN-SER remains competitive ($0.7812$) while operating with only 3 qubits, indicating that the MPS-structured circuit can provide effective correlation modeling without scaling the quantum width. On MDER, HQTN-SER does not match the highest reported accuracy; however, it uses 3 qubits and therefore serves as a reference point for resource-limited settings where increasing qubit count is not feasible. Overall, the comparison suggests that enforcing structure in the quantum circuit can support strong SER performance at small qubit counts, while higher peak accuracy may require substantially larger end-to-end models.

\begin{table}[htpbt]
\centering
\caption{Comparative accuracy and quantum footprint of HQTN-SER and representative quantum SER methods across SAVEE, MDER, and RAVDESS. Params denotes the \emph{total} trainable parameters (classical + quantum) as reported in the corresponding papers.}
\label{tab:comparative-results}
\small
\begin{tabularx}{\columnwidth}{@{}l l *{3}{>{\raggedleft\arraybackslash}X}@{}}
\hline
 & \textbf{Model} & \textbf{Acc.} & \textbf{Qubits} & \textbf{Params} \\
\hline
\multirow{3}{*}{\rotatebox{90}{\scriptsize \textbf{SAVEE}}}
& Qubit SW Deep-ESN \cite{soltani} & 0.7545 & 8--10 & 150k+ \\
& CDQKL \cite{chen2025consensus}             & 0.7875 & 5--6  & $\approx$10k \\
& \textbf{HQTN-SER}                          & \textbf{0.7812} & \textbf{3} & \textbf{3528} \\
\hline
\multirow{2}{*}{\rotatebox{90}{\scriptsize \textbf{MDER}}}
& HQNN \cite{mittal2025hybrid}               & 0.9450 & 6--8  & 200k+ \\
& \textbf{HQTN-SER}                          & \textbf{0.7351} & \textbf{3} & \textbf{3782} \\
\hline
\multirow{4}{*}{\rotatebox{90}{ \scriptsize \textbf{RAVDESS}}}
& Qubit SW Deep-ESN \cite{soltani} & 0.7983 & 8--10 & 150k+ \\
& CDQKL \cite{chen2025consensus}              & 0.7875 & 5--6  & $\approx$10k \\
& QMGU \cite{qu}                      & 0.7500 & 5--6 & $\approx$30k \\
& \textbf{HQTN-SER}                           & \textbf{0.8009} & \textbf{4} & \textbf{28364} \\
\hline
\end{tabularx}
\end{table}
\subsection{Hardware Analysis}

To evaluate the proposed model under realistic quantum-device noise, we perform a hardware-aware analysis on the MDER dataset. Although the proposed framework is evaluated on three datasets in the main experiments, we use MDER for this hardware-aware study because it provides a representative test case with stable baseline behavior and allows a focused assessment of whether the trained quantum representation remains robust when exposed to realistic device-level noise.

We re-evaluate the trained MDER model by replacing the noiseless \texttt{lightning.qubit} simulator with a noise-aware \texttt{qiskit.aer} execution setting. The quantum circuit weights are frozen after training; only the execution setting is changed, while the classical components and classifier remain unchanged.

We use \texttt{FakeMarrakesh} from Qiskit's fake provider as the hardware-aware backend. The corresponding fake backend is accessed through an \texttt{AerSimulator} and PennyLane's \texttt{qiskit.aer} device. All hardware-aware evaluations are performed using 1024 shots per circuit execution. Although this evaluation is not an execution on a live QPU, \texttt{FakeMarrakesh} is designed to emulate realistic IBM-style device behavior, including gate errors, readout errors, and shot-level sampling variability.

Table~\ref{tab:hardware} summarizes the hardware-aware evaluation. The noiseless \texttt{lightning.qubit} result reported earlier for MDER is 73.51\%. Under the \texttt{FakeMarrakesh} hardware-aware setting, the model achieves $73.45\% \pm 0.45$ pp across five evaluations, with individual runs ranging from 73.25\% to 74.25\%. This corresponds to a negligible shift of $-0.06$ percentage points relative to the noiseless setting.

\begin{table}[htpb]
\centering
\caption{Hardware-aware evaluation of MDER using \texttt{FakeMarrakesh}.}
\label{tab:hardware}
\footnotesize
\begin{tabular}{@{}ccc@{}}
\toprule
\textbf{Mean Acc.} & \textbf{Std.} & \textbf{Run Range} \\
\midrule
$73.45\%$ & $\pm 0.45$ pp & $73.25$--$74.25\%$ \\
\bottomrule
\end{tabular}
\end{table}
The close agreement between the noiseless and hardware-aware evaluations indicates that the trained MDER model preserves its predictive behavior under the tested device-noise setting. The small standard deviation across repeated evaluations also shows that the result is stable and not driven by a single favorable noisy run. Although one noisy run slightly exceeds the noiseless baseline, this should not be interpreted as noise improving the model. Rather, this variation is consistent with finite-shot sampling effects and stochastic noise-model behavior, especially since the difference remains within a narrow margin.

This stability can be linked to two structural properties of the MPS circuit. First, the shallow and locally connected circuit structure limits the number of quantum gates, reducing opportunities for noise accumulation across the circuit. Second, the use of expectation-value measurements as quantum features provides averaged observables rather than raw bitstring outputs, making the representation less sensitive to individual shot-level fluctuations. As a result, the compact MPS design, originally motivated by parameter efficiency and trainability, also exhibits passive robustness under the tested hardware-aware simulation setting.

\section{Conclusion}

This paper introduced HQTN-SER, a hybrid quantum-classical framework for speech emotion recognition that incorporates a matrix-product-state quantum tensor network circuit as a structured quantum module. The main goal was to examine whether MPS connectivity can provide a stable and resource-aware way to model correlations in speech features, rather than making claims of unconditional quantum advantage.
Experiments on RAVDESS, SAVEE, and MDER show that HQTN-SER achieves stable training behavior and competitive accuracy across benchmarks. The ablation study confirms that the full hybrid configuration is important in this setting: the classical-only variant is weaker, and the quantum-only variant performs poorly under limited qubit counts and shallow circuit depth. In contrast, HQTN-SER benefits from integrating classical embeddings with quantum measurement features, showing that the MPS-based circuit contributes useful structure when paired with learned classical representations.

The results also show consistent behavior across datasets with different speakers, languages, and recording conditions. In addition, the hardware-aware evaluation on MDER using \texttt{FakeMarrakesh} with 1024 shots shows that the trained model remains stable under realistic device-noise simulation, with only a negligible shift from the noiseless \texttt{lightning.qubit} result. This supports the practical relevance of the compact MPS design for near-term quantum machine learning studies, while execution on live QPUs remains an important next step.

Our proposed HQTN-SER provides an architecture-driven baseline for studying quantum tensor networks in SER, with an emphasis on reproducibility, compact circuit design, and operation under realistic quantum constraints. Future directions include richer feature encodings, improved measurement strategies, speaker-robust evaluation protocols, and hardware-executed inference as quantum devices and software tools continue to mature.
\section*{Acknowledgment}
 This work was supported in part by the NYUAD Center for Quantum and Topological Systems (CQTS), funded by Tamkeen under the NYUAD Research Institute grant CG008.

\begin{spacing}{0.983}
\bibliographystyle{IEEEtran}
 \footnotesize{
\bibliography{ref}}

@ARTICLE{911197,
  author={Cowie, R. and others},
    author2={Cowie, R. and Douglas-Cowie, E. and Tsapatsoulis, N. and Votsis, G. and Kollias, S. and Fellenz, W. and Taylor, J.G.},
  journal={IEEE Signal Processing Magazine}, 
  title={Emotion recognition in human-computer interaction}, 
  year={2001},
  volume={18},
  number={1},
  pages={32-80},
  doi={10.1109/79.911197}}

@article{ELAYADI2011572,
title = {Survey on speech emotion recognition: Features, classification schemes, and databases},
journal = {Pattern Recognition},
volume = {44},
number = {3},
pages = {572-587},
year = {2011},
issn = {0031-3203},
doi = {https://doi.org/10.1016/j.patcog.2010.09.020},
url1 = {https://www.sciencedirect.com/science/article/pii/S0031320310004619},
author2 = {Moataz {El Ayadi} and Mohamed S. Kamel and Fakhri Karray},
author={Moataz {El Ayadi} and others}
}

@article{article,
author = {Schuller, Björn},
year = {2018},
month1 = {04},
  volume={61},
  number={5},
  pages={90--99},
title = {Speech emotion recognition: Two decades in a nutshell, benchmarks, and ongoing trends},
journal = {Communications of the ACM},
doi = {10.1145/3129340}
}

@article{Schuld_2019,
   title={Quantum Machine Learning in Feature {Hilbert} Spaces},
  volume={122},
  number={4},
  pages={040504},
  year={2019},
  publisher={APS},
   ISSN={1079-7114},
   url1={http://dx.doi.org/10.1103/PhysRevLett.122.040504},
   DOI={10.1103/physrevlett.122.040504},
   journal={Physical Review Letters},
   author={Schuld, Maria and Killoran, Nathan},
   month1=feb }

@article{larocca2025barrenplateausvariationalquantum,
  title={Barren plateaus in variational quantum computing},
  author={Larocca, Martin and others},
  journal={Nature Reviews Physics},
  volume={7},
  number={4},
  pages={174--189},
  year={2025},
  publisher={Nature Publishing Group UK London}
}

@article{kard,
author = {Kardashin, Andrey and Uvarov, Aleksey and Biamonte, Jacob},
year = {2021},
month11 = {03},
pages = {586374},
title = {Quantum Machine Learning Tensor Network States},
volume = {8},
journal = {Frontiers in Physics},
doi = {10.3389/fphy.2020.586374}
}

@article{mao,
author = {Mao, Qirong and others},
year = {2014},
month11 = {12},
pages = {2203-2213},
title = {Learning Salient Features for Speech Emotion Recognition Using Convolutional Neural Networks},
volume = {16},
journal = {IEEE Transactions on Multimedia},
doi = {10.1109/TMM.2014.2360798}
}

@inproceedings{sperber2019selfattentionalmodelslatticeinputs,
  title={Self-attentional models for lattice inputs},
  author={Sperber, Matthias and others},
  booktitle={Proceedings of the 57th Annual Meeting of the Association for Computational Linguistics},
  pages={1185--1197},
  year={2019}
}

@inproceedings{radford2022robustspeechrecognitionlargescale,
  title={Robust speech recognition via large-scale weak supervision},
  author={Radford, Alec and others},
  booktitle={International conference on machine learning},
  pages={28492--28518},
  year={2023},
  organization={PMLR}
}

@article{latif,
  title={Survey of deep representation learning for speech emotion recognition},
  author={Latif, Siddique and others},
  journal={IEEE Transactions on Affective Computing},
  volume={14},
  number={2},
  pages={1634--1654},
  year={2021},
  publisher={IEEE}
}

@article{soltani,
  author={Soltani, Rebh and Benmohamed, Emna and Ltifi, Hela},
year = {2025},
month1 = {08},
  pages={77},
title = {Quantum-enhanced cortical deep echo state network for fast and accurate speech emotion recognition},
volume = {7},
journal = {Quantum Machine Intelligence},
doi = {10.1007/s42484-025-00304-1}
}

@article{qu,
author = {Qu, Zhiguo and others},
year = {2024},
month1 = {01},
  pages={6512--6523},
title = {{QFSM}: A Novel Quantum Federated Learning Algorithm for Speech Emotion Recognition With Minimal Gated Unit in 5G IoV},
  volume={9},
  number={10},
journal = {IEEE Transactions on Intelligent Vehicles},
doi = {10.1109/TIV.2024.3370398}
}

@article{Or_s_2014,
   title={A practical introduction to tensor networks: Matrix product states and projected entangled pair states},
   volume={349},
   ISSN={0003-4916},
   url1={http://dx.doi.org/10.1016/j.aop.2014.06.013},
   DOI={10.1016/j.aop.2014.06.013},
   journal={Annals of Physics},
   publisher={Elsevier BV},
   author={Orús, Román},
   year={2014},
   month1=oct, pages={117–158} }

@inproceedings{NIPS2016_5314b967,
 author = {Stoudenmire, Edwin and Schwab, David J},
 booktitle = {Advances in Neural Information Processing Systems},
 editor = {D. Lee and M. Sugiyama and U. Luxburg and I. Guyon and R. Garnett},
 publisher = {Curran Associates, Inc.},
 title = {Supervised Learning with Tensor Networks},
 url1 = {https://proceedings.neurips.cc/paper_files/paper/2016/file/5314b9674c86e3f9d1ba25ef9bb32895-Paper.pdf},
 volume = {29},
 year = {2016}
}

@article{ravdess,
author = {Livingstone, Steven and Russo, Frank},
year = {2018},
month1 = {05},
pages = {e0196391},
title = {The Ryerson Audio-Visual Database of Emotional Speech and Song {(RAVDESS)}: A dynamic, multimodal set of facial and vocal expressions in North American English},
volume = {13},
journal = {PLOS ONE},
doi = {10.1371/journal.pone.0196391}
}

@inproceedings{haq,
  title     = {{Audio-visual feature selection and reduction for emotion classification}},
  author    = {Sanaul Haq and Philip J. B. Jackson and James D. Edge},
  year      = {2008},
  booktitle = {{Auditory-Visual Speech Processing}},
  pages     = {185--190},
}

@misc{mderr,
doi = {10.21227/ev21-c430},
url1 = {https://dx.doi.org/10.21227/ev21-c430},
author = {{Mhamed-amine Soumiaa}},
publisher = {IEEE Dataport},
title = {Moroccan Dialect Emotion Recognition Dataset},
year = {2024} }

@inproceedings{chen2025consensus,
  title={Consensus-based distributed quantum kernel learning for speech recognition},
  author={Chen, Kuan-Cheng and Ma, Wenxuan and Xu, Xiaotian},
  booktitle={2025 IEEE International Conference on Acoustics, Speech, and Signal Processing Workshops (ICASSPW)},
  pages={1--5},
      doi = "10.1109/ICASSPW65056.2025.11011036",
  year={2025},
  organization={IEEE}
}

@article{alhussein2025speech,
  title={Speech emotion recognition in conversations using artificial intelligence: a systematic review and meta-analysis},
  author={Alhussein, Ghada and others},
  author2={Alhussein, Ghada and Ziogas, Ioannis and Saleem, Shiza and Hadjileontiadis, Leontios J},
  journal={Artificial Intelligence Review},
  volume={58},
  number={7},
  pages={198},
  year={2025},
  publisher={Springer}
}

@article{chowdhury2025speech,
  title={Speech emotion recognition with light weight deep neural ensemble model using hand crafted features},
  author={Chowdhury, Jaher Hassan and Ramanna, Sheela and Kotecha, Ketan},
  journal={Scientific Reports},
  volume={15},
  number={1},
  pages={11824},
  year={2025},
  publisher={Nature Publishing Group UK London}
}

@article{wu2025comprehensive,
  title={A comprehensive review of multimodal emotion recognition: Techniques, challenges, and future directions},
  author={Wu, You and others},
  journal={Biomimetics},
  volume={10},
  number={7},
  pages={418},
  year={2025},
  publisher={MDPI}
}

@article{biamonte2017quantum,
  title={Quantum machine learning},
  author={Biamonte, Jacob and others},
  author2={Biamonte, Jacob and Wittek, Peter and Pancotti, Nicola and Rebentrost, Patrick and Wiebe, Nathan and Lloyd, Seth},
  journal={Nature},
  volume={549},
  number={7671},
  pages={195--202},
  year={2017},
  publisher={Nature Publishing Group UK London}
}

@article{innan2024financial,
  title={Financial fraud detection using quantum graph neural networks},
  author={Innan, Nouhaila and others},
  author2={Innan, Nouhaila and Sawaika, Abhishek and Dhor, Ashim and Dutta, Siddhant and Thota, Sairupa and Gokal, Husayn and Patel, Nandan and Khan, Muhammad Al-Zafar and Theodonis, Ioannis and Bennai, Mohamed},
  journal={Quantum Machine Intelligence},
  volume={6},
  number={1},
  pages={7},
  year={2024},
  publisher={Springer}
}

@article{choudhary2026hqnn,
  title={{HQNN-FSP}: A hybrid classical-quantum neural network for regression-based financial stock market prediction},
  author={Choudhary, Prashant Kumar and others},
  journal={Quantum Machine Intelligence},
  volume={8},
  number={1},
  pages={55},
  year={2026},
  publisher={Springer}
}

@inproceedings{innan2025lep,
  title={Lep-qnn: Loan eligibility prediction using quantum neural networks},
  author={Innan, Nouhaila and Marchisio, Alberto and others},
  booktitle={2025 IEEE International Conference on Quantum Computing and Engineering (QCE)},
  volume={1},
  pages={1864--1872},
  year={2025},
  organization={IEEE}
}

@article{dave2025sentiqnf,
  title={Sentiqnf: A novel approach to sentiment analysis using quantum algorithms and neuro-fuzzy systems},
  author={Dave, Kshitij and others},
  journal={IEEE Transactions on Computational Social Systems},
  year={2025},
  publisher={IEEE}
}

@article{innan2024quantum,
  title={Quantum state tomography using quantum machine learning},
  author={Innan, Nouhaila and Siddiqui, Owais Ishtiaq and others},
  journal={Quantum Machine Intelligence},
  volume={6},
  number={1},
  pages={28},
  year={2024},
  publisher={Springer}
}

@inproceedings{innan2025quav,
  title={{QUAV}: Quantum-Assisted Path Planning and Optimization for UAV Navigation with Obstacle Avoidance},
  author={Innan, Nouhaila and Kashif, Muhammad and others},
  author2={Innan, Nouhaila and Kashif, Muhammad and Marchisio, Alberto and Gan, Yung-Sze and Barbaresco, Frederic and Shafique, Muhammad},
  booktitle={2025 IEEE International Conference on Quantum Artificial Intelligence (QAI)},
  pages={208--215},
  year={2025},
  organization={IEEE}
}

@article{innan2024financial1,
  title={Financial fraud detection: a comparative study of quantum machine learning models},
  author={Innan, Nouhaila and Khan, Muhammad Al-Zafar and Bennai, Mohamed},
  journal={International Journal of Quantum Information},
  volume={22},
  number={02},
  pages={2350044},
  year={2024},
  publisher={World Scientific}
}

@article{vyskubov2026scaling,
  title={Scaling Laws for Hybrid Quantum Neural Networks: Depth, Width, and Quantum-Centric Diagnostics},
  author={Vyskubov, Danil and others},
  journal={arXiv preprint arXiv:2604.06007},
  year={2026}
}

@article{balachandran2025advanced,
  title={Advanced speech emotion recognition utilizing optimized equivariant quantum convolutional neural network for accurate emotional state classification},
  author={Balachandran, G and others},
    author2={Balachandran, G and Ranjith, S and Jagan, GC and Chenthil, TR},
  journal={Knowledge-Based Systems},
  volume={316},
  pages={113414},
  year={2025},
  publisher={Elsevier}
}

@inproceedings{mittal2025hybrid,
  title={Hybrid quantum machine learning based human speech emotion recognition},
  author={Mittal, Sparsh and others},
  booktitle={2025 IEEE International Conference on Acoustics, Speech, and Signal Processing Workshops (ICASSPW)},
  pages={1--5},
  year={2025},
  organization={IEEE}
}

@article{norval2025quantum,
  title={Quantum ai in speech emotion recognition},
  author={Norval, Michael and Wang, Zenghui},
  journal={Entropy},
  volume={27},
  number={12},
  pages={1201},
  year={2025},
  publisher={MDPI}
}

@article{rajapakshe2025representation,
  title={Representation learning with parameterised quantum circuits for advancing speech emotion recognition},
  author={Rajapakshe, Thejan and others},
  journal={Scientific Reports},
  year={2025},
    volume={15},
  number={1},
  pages={44353},
  publisher={Nature Publishing Group UK London}
}

@inproceedings{krishna2024gesture,
  title={Gesture and emotion detection using quantum computing for enhanced recognition and analysis},
  author={Krishna, S Suraj Jai and others},
  booktitle={2024 International Conference on Expert Clouds and Applications (ICOECA)},
  pages={530--535},
  year={2024},
  organization={IEEE}
}

@inproceedings{kucharski2025survey,
  title={A Survey on Quantum Machine Learning in Speech Acoustics},
  author={Kucharski, Mateusz A},
  booktitle={2025 IEEE 25th International Symposium on Computational Intelligence and Informatics (CINTI)},
  pages={235--240},
  year={2025},
  organization={IEEE}
}

@article{golchha2025quantum,
  title={Quantum-based deep learning method for recognition of facial expressions},
  author={Golchha, Roopa and others},
  journal={Neural Computing and Applications},
  volume={37},
  number={16},
  pages={10163--10173},
  year={2025},
  publisher={Springer}
}

@article{barradas2026emotion,
  title={Emotion Recognition from Peripheral Physiological Signals: A Systematic Review of Trends, Challenges and Opportunities},
  author={Barradas, Isabel and others},
  journal={ACM Transactions on Interactive Intelligent Systems},
  volume={16},
  number={1},
  pages={1--41},
  year={2026},
  publisher={ACM New York, NY}
}

@inproceedings{solanki2025evaluating,
  title={Evaluating Multi-Layer Perceptron and Recurrent Neural Networks for Speech Emotion Recognition},
  author={Solanki, Satyam and others},
  booktitle={2025 3rd International Conference on Communication, Security, and Artificial Intelligence (ICCSAI)},
  volume={3},
  pages={349--354},
  year={2025},
  organization={IEEE}
}

@article{khalil2019speech,
  title={Speech emotion recognition using deep learning techniques: A review},
  author={Khalil, Ruhul Amin and others},
  journal={IEEE Access},
  volume={7},
  pages={117327--117345},
  year={2019},
  publisher={IEEE}
}

@article{issa2020speech,
  title={Speech emotion recognition with deep convolutional neural networks},
  author={Issa, Dias and Demirci, M Fatih and Yazici, Adnan},
  journal={Biomedical Signal Processing and Control},
  volume={59},
  pages={101894},
  year={2020},
  publisher={Elsevier}
}

@article{swain2018databases,
  title={Databases, features and classifiers for speech emotion recognition: a review},
  author={Swain, Monorama and Routray, Aurobinda and Kabisatpathy, Prithviraj},
  journal={International Journal of Speech Technology},
  volume={21},
  number={1},
  pages={93--120},
  year={2018},
  publisher={Springer}
}

@article{hong2025review,
  title={A review on quantum machine learning in applied systems and engineering},
  author={Hong, Ying-Yi and Lopez, Dylan Josh Domingo},
  journal={IEEE Access},
  year={2025},
  publisher={IEEE}
}
\end{spacing}
\end{document}